\documentstyle[preprint,aps]{revtex}
\tighten
\draft
\widetext
\input epsf
\preprint{YITP-SB-02-03, NSF-ITP-02-11}
\bigskip
\bigskip

\begin{document}
\title{A Remark on Smoothing Out Higher Codimension Branes}
\medskip
\author{Olindo Corradini$^1$\footnote{E-mail: olindo@insti.physics.sunysb.edu},
Alberto Iglesias$^1$\footnote{E-mail: iglesias@insti.physics.sunysb.edu},
Zurab Kakushadze$^1$\footnote{E-mail: zurab@insti.physics.sunysb.edu} and Peter
Langfelder$^{1,2}$\footnote{E-mail: plangfel@insti.physics.sunysb.edu}}
\bigskip
\address{$^1$C.N. Yang Institute for Theoretical Physics\\ 
State University of New York, Stony Brook, NY 11794\\
$^2$Institute for Theoretical Physics, University of California, 
Santa Barbara, CA 93106}
\date{January 24, 2002}
\bigskip
\medskip
\maketitle

\begin{abstract} 
{}We discuss some issues arising in studying (linearized) 
gravity on non-BPS higher
codimension branes in an infinite-volume bulk. In particular, such backgrounds
are badly singular for codimension-3 and higher $\delta$-function-like branes
with non-zero tension.
As we discuss in this note, non-trivial issues arise in smoothing out such
singularities. Thus, adding higher curvature terms might be necessary in this 
context. 
\end{abstract}
\pacs{}

\section{Introduction and Summary}

{}As was originally proposed in \cite{DG}, 
one can reproduce four-dimensional gravity
on a 3-brane in 6 or higher dimensional infinite-volume bulk if one includes
an induced Einstein-Hilbert term on the brane. Gravity then is almost
completely localized on the brane with only ultra-light modes penetrating into
the bulk, so that gravity is four-dimensional at distance scales up to
an ultra-large cross-over scale $r_c$ (beyond which gravity becomes higher 
dimensional), which can be larger than the present Hubble size.
In particular, this is the case for codimension-2 and higher 
tensionless branes \cite{DG} as well as for codimension-2 non-zero tension
branes \cite{CIKL}.

{}A careful analysis of linearized gravity in such backgrounds requires
smoothing out higher codimension singularities \cite{DG,orient}. This is
already the case for tensionless branes, where the background is non-singular 
(in fact, it is flat), but the graviton propagator is singular. 
In the case of non-zero tension branes the situation is even more complicated
as in the case of $\delta$-function-like branes
the background itself becomes singular (for phenomenologically interesting
non-BPS branes on which we focus in this paper). More precisely, in the
codimension-2 case the singularity is very mild as in the extra 2 dimensions
the background has the form of a wedge with a deficit angle, so the 
singularity is a simple conical one \cite{CIKL}. 
As was discussed in \cite{orient}, this
singularity can be consistently smoothed out.
In this case the gravity on the
brane was analyzed in \cite{CIKL,orient}, 
where it was found that the behavior of
gravity is essentially unmodified compared with the tensionless brane cases.

{}In codimension-3 and higher cases the singularities are more severe.
The purpose of this note is to study these backgrounds using a smoothing out 
procedure discussed in \cite{orient}. This procedure goes as follows. Consider
a codimension-$d$ $\delta$-function-like source brane in $D$-dimensional bulk.
Let the bulk action simply be the $D$-dimensional Einstein-Hilbert action,
while on the brane we have the induced $(D-d)$-dimensional Einstein-Hilbert 
term as well as the cosmological term corresponding to the brane tension.
As we have already mentioned, the background in this case is singular 
\cite{Gregory}. One way to smooth out such a singularity is to replace the
$(D-d)$-dimensional world-volume of the brane by its product with a 
$d$-dimensional ball ${\bf B}_d$ of some non-zero radius $\epsilon$. 
As was pointed out in \cite{orient}, in this case already for a tensionless 
brane the gravitational modes on the brane contain an infinite tower
of tachyonic modes. This can be circumvented by considering a partial smoothing
out where one replaces the  
$(D-d)$-dimensional world-volume of the brane by its product with a 
$(d-1)$-sphere ${\bf S}^{d-1}$ of radius $\epsilon$ \cite{orient}. 
As was pointed out in \cite{orient}, this suffices for 
smoothing out higher codimension singularities in the graviton propagator
as in the codimension-1 case the propagator is non-singular \cite{DGP}. 
Moreover,
in the case of tensionless branes as well as in the case of a non-zero
tension codimension-2 brane we then have only one tachyonic mode which is 
expected to be an
artifact of not including non-local operators on the brane. The question
we would like to address here is whether this smoothing out procedure can also
cure singularities of the background itself in the case of 
codimension-3 and higher non-zero tension branes.

{}We find that there are no non-singular solutions of the aforementioned 
type. One possibility here could be to add higher curvature terms which might
help cure these singularities. In particular, as was recently discussed in
\cite{curvature}, higher curvature terms are expected to 
smooth out higher codimension singularities in the graviton propagator. 

\section{Setup}

{}The brane world model we study in this paper is described 
by the following action:
\begin{equation}\label{action}
 S={\widetilde M}^{D-3}_P \int_{\Sigma} d^{D-1}x~
 \sqrt{-{\widetilde G}}\left[{\widetilde R}-{\widetilde\Lambda}\right]
 +{M}^{D-2}_P \int d^{D}x~
 \sqrt{-{G}}~{R}~.
\end{equation}
Here $M_P$ is the (reduced) $D-$dimensional Planck mass, while 
${\widetilde M}_P$
is the (reduced) $(D-1)$-dimensional Planck mass; $\Sigma$ is a 
source brane, whose geometry is given by 
the product ${\bf R}^{D-d-1,1}\times {\bf S}^{d-1}_\epsilon$, where 
${\bf R}^{D-d-1,1}$ is the $(D-d)$-dimensional Minkowski space, and 
${\bf S}^{d-1}_\epsilon$ is a $(d-1)$-sphere of radius $\epsilon$ (in the
following we will assume that $d\geq 3$).
The quantity ${\widetilde M}^{D-3}_P {\widetilde \Lambda}$ plays the role of 
the tension of the brane $\Sigma$. Also, 
\begin{equation}
 {\widetilde G}_{mn}\equiv
 {\delta_m}^M{\delta_n}^N G_{MN}\Big|_\Sigma~,
\end{equation} 
where $x^m$ are the $(D-1)$ coordinates along the brane (the $D$-dimensional
coordinates are given by $x^M=(x^m,r)$, where $r\geq 0$ is a non-compact
radial coordinate transverse to the brane, and the signature of the 
$D$-dimensional metric is $(-,+,\dots,+)$); finally, the $(D-1)$-dimensional
Ricci scalar ${\widetilde R}$ is constructed from the $(D-1)$-dimensional
metric ${\widetilde G}_{\mu\nu}$. In the following we will use the notation
$x^i=(x^\alpha,r)$, where $x^\alpha$ are the $(d-1)$ angular 
coordinates on the sphere.
Moreover, the metric for the coordinates $x^i$ will be (conformally) flat:
\begin{equation}
 \delta_{ij}~dx^i dx^j=dr^2+r^2\gamma_{\alpha\beta}~dx^\alpha dx^\beta~,
\end{equation}
where $\gamma_{\alpha\beta}$ is the metric on a unit $(d-1)$-sphere.
Also, we will
denote the $(D-d)$ Minkowski coordinates on ${\bf R}^{D-d-1,1}$ via $x^\mu$
(note that $x^m=(x^\mu,x^\alpha)$).

{}The equations of motion read
\begin{eqnarray}
 &&R_{MN}-\frac{1}{2} G_{MN}R+\displaystyle{
 {\sqrt{-{\widetilde G}}\over\sqrt{-G}}}
 {\delta_M}^m {\delta_N}^n \left[{\widetilde R}_{mn}-{1\over 2} 
 {\widetilde G}_{mn}\left({\widetilde R}-
{\widetilde \Lambda}\right)\right]  {\widetilde L}~ \delta(r-\epsilon)=0~,
 \label{EoM4}
\end{eqnarray}
where 
\begin{equation}
 {\widetilde L}\equiv {\widetilde M}_P^{D-3}/M_P^{D-2}~.
\end{equation}
To solve these equations, let us use the following ansatz for the 
background metric:
\begin{equation}
 ds^2=\exp(2A)~\eta_{\mu\nu}~dx^\mu dx^\nu+\exp(2B)~\delta_{ij}~dx^i dx^j~,
\end{equation}
where $A$ and $B$ are
functions of $r$ but are independent of $x^\mu$ and $x^\alpha$ (that is, we are
looking for solutions that are radially symmetric in the extra dimensions). 
We then have (here prime
denotes derivative w.r.t. $r$):
\begin{eqnarray}
 &&{\widetilde R}_{\mu\nu}=0~,\\
 &&{\widetilde R}_{\alpha\beta}=\lambda ~{\widetilde G}_{\alpha\beta}~,\\
 &&{\widetilde R}=(d-1)~\lambda~,\\
 &&R_{\mu\nu}=-\eta_{\mu\nu}e^{2(A-B)}
 \left[A^{\prime\prime} +(d-1){1\over r} A^\prime+
 (D-d)(A^\prime)^2 +(d-2)A^\prime B^\prime\right]~,\\
 &&R_{rr}=-(d-1)\left[B^{\prime\prime}+ {1\over r} B^\prime\right]
 +(D-d)\left[A^\prime B^\prime-
 (A^\prime)^2 -A^{\prime\prime}\right]~,\\
 &&R_{\alpha\beta}=-r^2 \gamma_{\alpha\beta}
 \left[B^{\prime\prime}+(2d-3){1\over r}B^\prime +(d-2)(B^\prime)^2+
 (D-d)A^\prime B^\prime + (D-d){1\over r}A^\prime\right]~,\\
 &&R=-e^{-2B}\Big[2(d-1)B^{\prime\prime} + 2(d-1)^2{1\over r} B^\prime +
 2(D-d)A^{\prime\prime} + 2(D-d)(d-1){1\over r} A^\prime 
 +\nonumber\\
 &&(d-1)(d-2)(B^\prime)^2 +(D-d)(D-d+1)(A^\prime)^2
 +2(D-d)(d-2)A^\prime B^\prime\Big]~, 
\end{eqnarray}
where
\begin{equation}
 \lambda\equiv {{d-2}\over \epsilon^2}e^{-2B(\epsilon)}~.
\end{equation}
The equations of motion then read:
\begin{eqnarray}\label{AB1}
 &&(D-d)\left[{1\over 2}(D-d-1)(A^\prime)^2+(d-1){1\over r}A^\prime+
 (d-1)A^\prime B^\prime\right]+\nonumber\\
 &&(d-1)(d-2)\left[{1\over 2}(B^\prime)^2+{1\over r}B^\prime\right]
 =0~,\\
 &&(D-d)\left[A^{\prime\prime}+
 {1\over 2}(D-d+1)(A^\prime)^2+(d-2){1\over r}A^\prime+(d-3)
 A^\prime B^\prime\right]+\nonumber\\
 &&(d-2)\left[B^{\prime\prime}+{1\over 2}(d-3)(B^\prime)^2+(d-2){1\over r}
 B^\prime\right]+{1\over 2}e^{B}\left[{\widetilde\Lambda}-(d-3)\lambda\right]~
 {\widetilde L}~\delta(r-\epsilon)=0~,\label{AB2}\\
 &&(D-d-1)\left[A^{\prime\prime}+{1\over 2}(D-d)(A^\prime)^2+(d-1){1\over r}
 A^\prime+(d-2)A^\prime B^\prime\right]+\nonumber\\
 &&(d-1)\left[B^{\prime\prime}+{1\over 2}(d-2)(B^\prime)^2+(d-1){1\over r}
 B^\prime\right]+{1\over 2}e^{B}\left[{\widetilde\Lambda}-(d-1)\lambda\right]~
 {\widetilde L}~\delta(r-\epsilon)=0~.\label{AB3}
\end{eqnarray}
Here the third equation is the $(\mu\nu)$ equation, the second equation is the
$(\alpha\beta)$ equation, while the first equation is the $(rr)$ equation.
Note that the latter equation 
does not contain second derivatives of $A$ and $B$.
The solution for $B^\prime$ is given by (we have chosen the plus root,
which corresponds to solutions with infinite-volume extra space):
\begin{equation}\label{root}
 B^\prime=-{1\over r}-{{D-d}\over{d-2}} A^\prime +
 \sqrt{{1\over r^2}+{1\over \kappa^2} (A^\prime)^2 }~,
\end{equation}
where we have introduced the notation
\begin{equation}
 {1\over \kappa^2}\equiv {(D-d)(D-2)\over (d-1)(d-2)^2}
\end{equation} 
to simplify the subsequent equations.

\section{The Background}

{}We can solve the above equations of motion as follows. First, consider
the difference of (\ref{AB1}) and (\ref{AB2}):
\begin{eqnarray}
 &&(D-d)\left[2A^\prime B^\prime-(A^\prime)^2-A^{\prime\prime}+
 {1\over r}A^\prime\right]+\nonumber\\
 &&(d-2)\left[(B^\prime)^2-B^{\prime\prime}+{1\over r} B^\prime\right]
 -{1\over 2}e^{B}\left[{\widetilde\Lambda}-(d-3)\lambda\right]~
 {\widetilde L}~\delta(r-\epsilon)=0~.\label{alt}
\end{eqnarray}
Plugging (\ref{root}) into this equation we obtain the following equation:
\begin{eqnarray}
 &&{A^\prime\left[rA^{\prime\prime}+A^\prime\right]\over
 \sqrt{\kappa^2+r^2 (A^\prime)^2 }} + {(d-2)\over \kappa} (A^\prime)^2 +
 {\kappa\over 2(d-2)}e^{B}\left[{\widetilde\Lambda}-(d-3)\lambda\right]~
 {\widetilde L}~\delta(r-\epsilon)=0~.
\end{eqnarray}
This equation can then be integrated. Thus, let
\begin{equation}
 Q\equiv {1\over\kappa} rA^\prime~.
\end{equation}
Then we have
\begin{equation}\label{Q}
 {Q Q^\prime\over\sqrt{1+Q^2}}+(d-2){1\over r}Q^2+{\epsilon\over 2(d-2)}
 e^{B}\left[{\widetilde\Lambda}-(d-3)\lambda\right]~
 {\widetilde L}~\delta(r-\epsilon)=0~.
\end{equation}
Here we are interested in non-singular
solutions such that $A$ and $B$ are constant
for $r<\epsilon$, and asymptote to some finite values as $r\rightarrow\infty$.
The corresponding solution for $Q(r)$ is given by (here $\theta(x)$ is the
Heavyside step-function):
\begin{eqnarray}
 &&Q(r)={2f(r)\over {1-f^2(r)}}~\theta(r-\epsilon)~,
\end{eqnarray}
where
\begin{equation}
 f(r)\equiv \left({r_*\over r}\right)^{d-2}~,
\end{equation}
and $r_*$ is the integration constant. Note that due to the discontinuity at 
$r=\epsilon$ we have the following matching condition:
\begin{equation}\label{match1}
 (d-2) {2f^2(\epsilon)\over{1-f^2(\epsilon)}}+{\epsilon{\widetilde L}\over 2}
 e^{B(\epsilon)}\left[{\widetilde\Lambda}-(d-3)\lambda\right]=0~.
\end{equation}
Note that, as it should be, this matching condition is the same as the one
that follows from equation (\ref{alt}).

{}Next, we solve for $A$ and $B$:
\begin{eqnarray}
 &&A(r)=A(\epsilon)~,~~~r\leq\epsilon~,\\
 &&A(r)=A_\infty -{\kappa\over{d-2}}~
 \ln\left({{1+f(r)}\over{1-f(r)}}\right)~,~~~r>\epsilon~,\\
 &&B(r)=B(\epsilon)~,~~~r\leq\epsilon~,\\
 &&B(r)=B_\infty-{{D-d}\over{d-2}}\left(A(r)-A_\infty\right)+{1\over{d-2}}~
 \ln\left(1-f^2(r)\right)~,~~~r>\epsilon~,
\end{eqnarray}
where $A_\infty$ and $B_\infty$ are the asymptotic values of 
$A(r)$ respectively
$B(r)$ as $r\rightarrow \infty$.

{}To complete our task here, we must check that (\ref{AB3}) is also satisfied. 
In fact, it is more convenient to check the difference of (\ref{AB3}) and 
(\ref{AB1}):
\begin{eqnarray}
 &&(D-d-1)A^{\prime\prime}-(d-1){1\over r}A^\prime-(D-2)A^\prime B^\prime
 +\nonumber\\
 &&(d-1)\left[B^{\prime\prime}+{1\over r}B^\prime\right]+
 {1\over 2}e^{B}\left[{\widetilde\Lambda}-(d-1)\lambda\right]~
 {\widetilde L}~\delta(r-\epsilon)=0~.
\end{eqnarray}
By direct examination it is not difficult to check that this equation is 
indeed satisfied subject to the following matching condition:
\begin{equation}\label{match2}
 (d-1){2f^2(\epsilon)\over{1-f^2(\epsilon)}}-{{D-2}\over {d-2}}
 {2\kappa f(\epsilon)\over{1-f^2(\epsilon)}} +
 {\epsilon{\widetilde L}\over 2}
 e^{B(\epsilon)}\left[{\widetilde\Lambda}-(d-1)\lambda\right]=0~.
\end{equation}
We can rewrite the matching conditions (\ref{match1}) and (\ref{match2})
as follows:
\begin{eqnarray}
 &&(d-2) {2f^2(\epsilon)\over{1-f^2(\epsilon)}}+{\epsilon{\widetilde L}\over 2}
 e^{B(\epsilon)}\left[{\widetilde\Lambda}-(d-3)\lambda\right]=0~,\\
 &&(D-2) {2\kappa f(\epsilon)
 \over{1-f^2(\epsilon)}}+{\epsilon{\widetilde L}\over 2}
 e^{B(\epsilon)}\left[{\widetilde\Lambda}+(d-1)\lambda\right]=0~.
\end{eqnarray}
Let us study possible solutions to these matching conditions.

{}Thus, let us see if there are non-singular solutions with 
$r_*<\epsilon$ for
which $f(\epsilon)<1$. Then these matching conditions can only be satisfied
if ${\widetilde\Lambda}<0$. On the other hand, the parameter $\lambda$ is
positive. Let
\begin{equation}
 {\widetilde \Lambda}\equiv -\gamma\lambda~,
\end{equation}
where $\gamma$ is a positive parameter. Then we have:
\begin{equation}
 f(\epsilon)=\rho~{{\gamma+d-3}\over {\gamma-d+1}}~,
\end{equation}
where
\begin{equation}
 \rho\equiv \kappa {{D-2}\over{d-2}} =\sqrt{(d-1)(D-2)\over{D-d}}~.
\end{equation}
Note that we must have $\gamma>d-1$. On the other hand, 
note that for $d\geq 3$ we always have $\rho>\sqrt{2}$. This then implies that
there is no solution with $f(\epsilon)<1$, that is, there is no solution with
$r_*<\epsilon$. In other words, there are no non-singular solutions of this
type. That is, there are no solutions with non-zero brane tension (for
vanishing brane tension smooth solutions were discussed in \cite{orient}).

{}Let us end with the following remark. Note that in the action (\ref{action})
we included not only the $(D-d)$-dimensional Einstein-Hilbert term for the
Minkowski part ${\bf R}^{D-d-1,1}$ but also for the sphere part. It is not
difficult to see that dropping (or rescaling) 
the latter does not change any conclusions.

\acknowledgments

{}This work was supported in part by the National Science Foundation grants
PHY-0071400 and PHY-9907949 and an Alfred P. Sloan Fellowship.
P.L. would like to thank the Institute for Theoretical Physics 
at the University of California at Santa 
Barbara for their hospitality while parts of this work were completed.
Z.K. would also like to thank Albert and Ribena Yu for financial support.

\end{document}